\documentclass[11pt,twoside]{article}
\usepackage{asp2010}
\usepackage[T1]{fontenc}

\resetcounters

\begin{document}

\title{Evolution of the IVOA Characterisation Data Model }
\author{ Bonnarel Fran\c{c}ois$^1$, Louys Mireille,$^{1,2}$ and  Chilingarian Igor$^3$ }
\affil{$^1$CDS, Observatoire astronomique de Strasbourg, 11 Rue de l'Universit$\acute{e}$, 67000 Strasbourg, France}
\affil{$^2$Icube Laboratory, UDS, 67400 Illkirch, France}
\affil{$^3$SAO, Harvard Smithsonian Center for Astrophysics, Cambridge, USA}
\begin{abstract}
The Characterisation data model \citep{2011arXiv1111.2281L} is a standard of the International Virtual Observatory Alliance (IVOA) that describes observational datasets in the multi-dimensional parameter space. Defining three properties: coverage, resolution, and sampling along different physical axes (e.g. spatial, spectral, flux) with variable level of details for the description, this model has been used in several IVOA contexts: Simple Spectral Access Protocol, Spectrum Data Model, ObsTAP (Table Access Protocol for the Core Components of the Observation Data Model \citep{2011arXiv1111.1758L}. Here we propose a  new version  which addresses more completely the most detailed level of description (level 4) dealing with variation maps of coverage, resolution, and sampling. It also introduces new specific axes in order to cover various photometric measurements, velocity and polarimetry. Special care is given for composed data sets. These improvements and add-ons follow the evolution of needs expressed to the uptake of VO tools in various observation domains for data discovery but also for data analysis requirements. This is also introduced to tackle use-cases designed to analyse scientifically datasets in the V0 context together with calibration/provenance information.

\end{abstract}

\section{Introduction}

The Observation data model project appeared at the first Data Model forum held at the May 2003 IVOA meeting in Cambridge,UK. Rapidly some
main classes appeared to be necessary to organise the metadata: Dataset
or Observation, Identification, Physical Characterisation, Provenance (either
instrumental or software) and Curation. A description of the early stages of
this development can be found in the Observation DM IVOA Note \citep{OBSNOTE}. In
parallel, an effort dedicated to spectra was lead by the DM Working group.
The Spectrum data model represents all necessary metadata for one specific
type of observational data: simple spectra. For the overall Observation Data
Model, the physical characterisation has been identified to be on first
 priority already in 2004. It was completed as an IVOA recommendation after 4
years of discussion which included computer scientists, astronomers and data
providers under the lead of J. McDowell.
In theory full characterisation should be done by  describing the  transformation of incoming radiation into the actual measurements.     In practice the effects of this observation process can be split into various levels and pieces of information. The IVOA Characterisation data model organises metadata as a 3D matrix
spanning independently the various physical axes: spatial, spectral, time,
flux or whatever observable quantity (1st dimension), and describing for
each of these axes 3 kind of properties, namely coverage (or sensitivity), resolution and sampling. The third dimension is the level of description of these items, from coarse
average or typical values of parameters down to variation maps. This scheme allows to support selection of data sets for data discovery as well as data analysis.

\section{Characterisation DM usage for Data Discovery. New use cases related to VO applications.}
    While the Characterisation data model was setting up a logical framework to describe the properties/features of each observation in the VO, it was lacking a simple DAL access protocol, except in the spectral case \citep{SSA}. Using the emerging TAP/SCHEMA
framework, ObsTAP services fill the gap, using  Characterisation properties and providing other data model fields to describe any kind of observation product.  The emerging datalink protocol \citep{DATALINK} should help to  access to full characterisation and observation metadata. 
     Various applications make use of Characterisation instance documents attached to data products. 
     \begin{itemize}
	\item CAMEA \citep{CAMEA}is a visualizer/editor of Characterization classes  for individual data products.    
    
	\item 
The CDS Aladin SED Plugin \citep{SEDPlug} extracts various information from the characterization metadata associated to images in order to plot SED of photometric extractions  in extended sources and helps to analyze them .
  \item The Aida workflow interface \citep{2009mavo.proc...95L}    uses the Characterisation XML instances of data products to check and validate a sequence of steps and execute it.  
\end{itemize}
\begin{figure}[]
	\begin{center}
		\includegraphics[height=8cm]{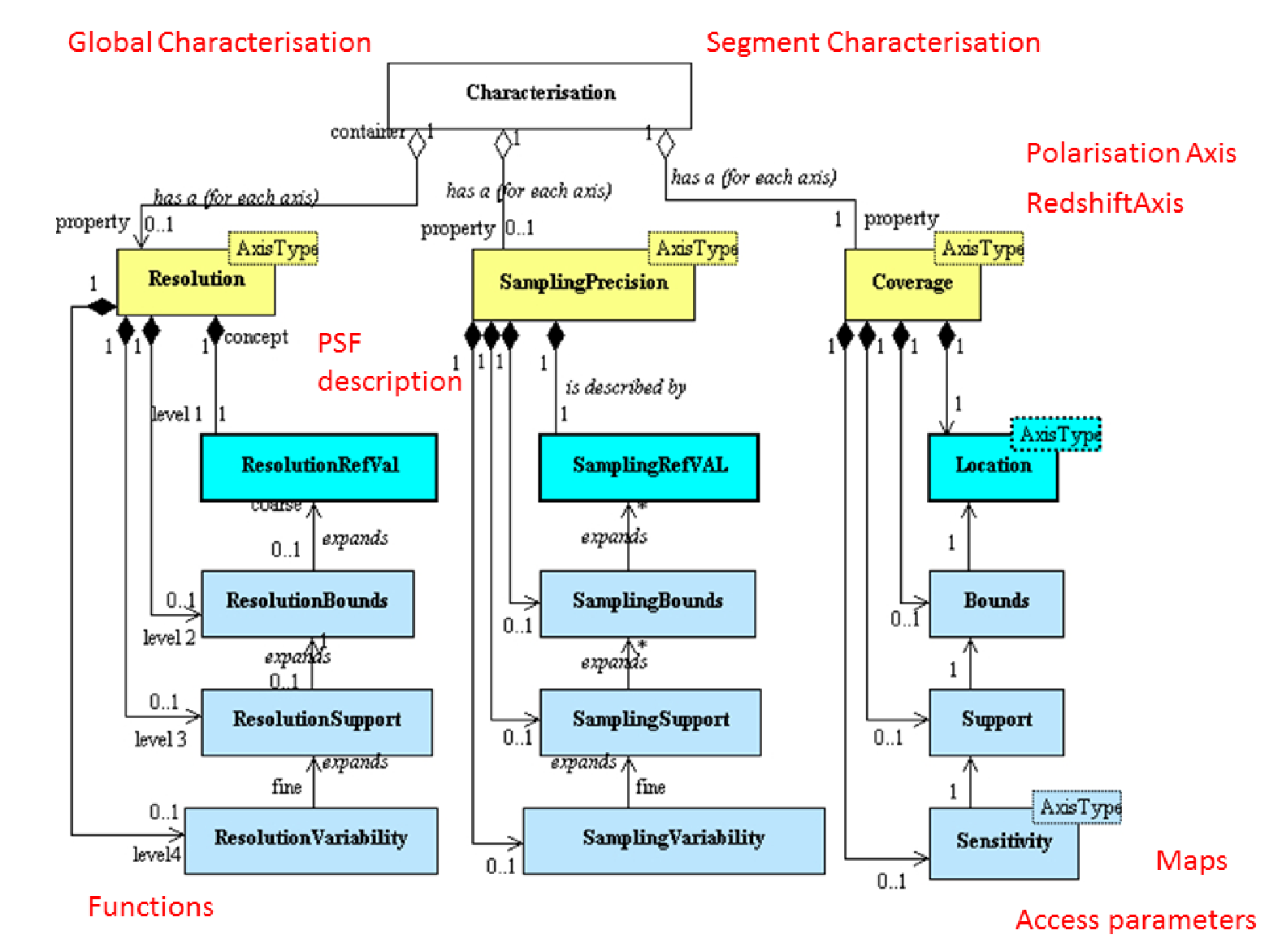}
	\end{center}
\caption{New features for the Characterisation v2 data model}
	\label{fig:Fig.1}
\end{figure}

    But in all these attempts, the actual full variability of characterisation properties is currently missing. 

      Comparison between observations and phenomenological or theoretical models requires  knowledge of the local transformation effects applied on the signal during the observation process: this is what Characterisation level 4 is for.
     These are two use cases  which could require  level 4 Characterisation:
\begin{itemize}
 \item     Deconvolution of point source or extended sources signal from a variable PSF for separation of stars in a crowded field, or better measurement of galaxy parameters.
\item Fitting a theoretical spectrum against observed spectral data taking into account the varying spectral resolution which has been pushed towards a real VO Web service using features described in this paper \citep{chil_adassxxi}  
\end{itemize}
      The main issue for these applications is to provide ways  to access to the description of variation maps like for example sensitivity maps,  
      PSF, LSF, variable FWHM of the response functions.

We also address the case of an observation product actually made of several sub-observations with subparts showing consistent supports in coverage resolution and sampling.  Most common examples for that are: CCD mosaics, echelle spectra and raw IFU data. The characterisation of such complex or composed observations clearly need two levels of description: a common one for the whole observation as well as individual characterisation of each segment (see section 4).

Specialized axes in the model take into account the peculiarities of the various physical parameters. Spatial, spectral, Time, Observable and Flux axes are common concepts already defined in Char DM 1.13. A Velocity (or redshift) axis is required
 to describe spectral lines observations,  such as radio cubes or H$\alpha$ cubes where for each position in the spatial field of view the third axis gives a measurement of a Doppler shift from the rest wavelength of the observed spectral line.
Polarization is a special parameter in this that whatever system we use to describe it, there is a very small number of possible values (such as Stokes parameters). This axis is actually discrete.

\section{New features for Characterisation DM version 2}

Figure 1 shows a diagram of the characterisation model with red fonts for all the new features, namely: description of a PSF, two new roles for Characterisation of complex data: global characterisation and segment characterisation, new specialized axes such as Velocity axis and Polarisation axis with list of discrete  set of values. and eventually descriptions of the variations maps at level 4.

Figure 2 describes the UML diagram illustrating the various possibilities for  describing a variation map or a PSF in Characterisation 2.0. Functional descriptions can be given using a MathML type of format or simple \textit{C}-like expressions. An alternative can be obtained by using a limited set of statistical moments. Last but not least an extended access formalism allows to describe internal structure of complex datasets in order to allow variation maps stored as bitmaps in some files at a specified URL. 

Composed  Observations can be described by two different flavors of characterisation classes: the Global characterisation well suited to data discovery retains the common features of the whole observations, while the segment ones, more adapted to data analysis describe each sub-observation in detail. New Characterisation class methods allow to navigate from Global Characterisation to segments ones and vice versa.\\ \\
Characterisation Serialization will also be updated by publishing soon a revised XML schema and revised list of utypes.

\begin{figure}[]
	\begin{center}
		\includegraphics[height=6cm]{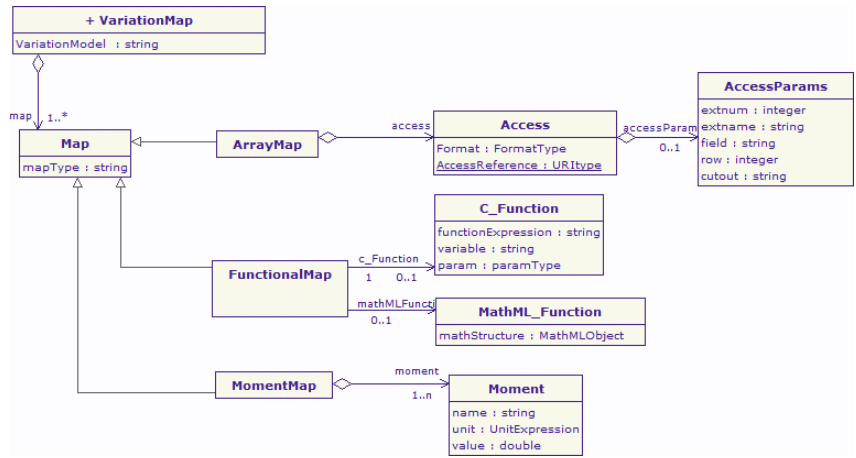}
	\end{center}
\caption{Details of level 4 description}
	\label{fig:Fig.2}
\end{figure}

\bibliographystyle{asp2010}

\bibliography{P016}

\end{document}